\documentclass[pra,twocolumn,showpacs,preprintnumbers,amsmath,amssymb]{revtex4}
\usepackage{graphicx}		

\begin{document}
\title{Rogue decoherence in the formation of a macroscopic atom-molecule superposition}
\author{Olavi Dannenberg}
\email{olavi.dannenberg@helsinki.fi}
\affiliation{Helsinki Institute of Physics, PL 64, FIN--00014 Helsingin yliopisto, Finland}
\affiliation{Department of Physics, University of Turku, FIN--20014 Turun yliopisto, Finland}
\author{Matt Mackie}
\affiliation {Department of Physics, Temple University, Philadelphia, PA 19122}
\affiliation{Department of Physics, University of Connecticut, Storrs, CT 06268}
\date{\today}

\begin{abstract}
We theoretically examine two-color photoassociation of a Bose-Einstein
condensate, focusing on the role of rogue decoherence in the formation
of macroscopic atom-molecule superpositions. Rogue dissociation occurs
when two zero-momentum condensate atoms are photoassociated into a
molecule, which then dissociates into a pair of atoms of
equal-and-opposite momentum, instead of dissociating back to the
zero-momentum condensate. As a source of decoherence that may damp
quantum correlations in the condensates, rogue dissociation is an
obstacle to the formation of a macroscopic atom-molecule superposition.
We study rogue decoherence in a setup which, without
decoherence, yields a macroscopic atom-molecule superposition, and
find that the most favorable conditions for said superposition
are a density $\rho\sim10^{12}\,{\rm cm}^{-3}$ and temperature $T\sim10^{-10}\,{\rm K}$. 
\end{abstract}
\maketitle

\section{Introduction}
Since the famous thought experiment by Schr\"odinger \cite{sc35}, macroscopic superpositions have
been a
part of quantum folklore that have annoyed specialists of the field. Schr\"odinger considered a cat
that was sealed into a steel chamber with a diabolic device that consists of a small amount of radioactive
material and a Geiger counter that was attached to a hammer device. If one radioactive nucleus would decay,
then the Geiger counter would trigger the hammer device, and the hammer device would scrap a small bottle
of Prussic acid which kills the cat. If the problem is
studied strictly quantum mechanically, from outside the steel chamber, one finds that after a while the cat
would be in a superposition state of alive and dead. If the cat would be in the statistical
mixture of being alive or dead, the situation would not differ much from, e.g., tossing a coin with the
probability of 0.5 for heads or tails. In a classical model, we do not yet {\it know} the outcome, but the
state of the cat (coin) {\it is} either alive or dead (heads or tails), even before the measurement. The
quantum cat (coin) behaves differently, with laws that allow superpositions between the outcome alive and
outcome dead (heads and tails). Hence, rather than certain alive or certain dead (heads or tails)
before the measurement, there must be quantum correlations between the the two outcomes.

For Schr\"odinger, the reason for putting the cat in the experimental spotlight was obvious. The absurd
outcome of the experiment illustrated that, if the outcome was a result of treating everything--including
the cat--strictly quantum mechanically, something was wrong, if not in the theory itself, then at least in
the understanding quantum mechanics. As previous decoherence studies have shown
\cite{zu82,cl83,uz89,zu91,hpz92,alzp95,br96}, one should be wary of too many oversimplifications and
approximations. The environment of the cat should also be modelled, not only a point-like cat in a vacuum.
The interaction between the cat and the environment induces decoherence that damps quantum correlations out
of the cat, resulting in a classical statistical mixture of living and dead cat. In other words, the cat is
too macroscopic to be in a superposition state. But what, then, is the upper limit of macroscopicity for a
superposition state?

Recent studies \cite{ru98, ci98, cs99, dalvit00, cms01, hm05} have pointed out that the macroscopicity
question can perhaps be addressed with Bose-Einstein condensates. References \cite{ru98, ci98, cs99,
dalvit00} consider superpositions of two-component atomic condensates, while dynamical creation of a
macroscopic superposition between atomic and molecular condensates via photoassociation \cite{cms01} or
Feshbach resonance \cite{hm05} has also been studied. The work herein is a follow-up study
of \cite{cms01}, one that includes dissociative damping of quantum correlations. In particular, we consider
a macroscopic atom-molecule superposition created in a two step process: first, a joint
atom-molecule condensate is created from an initial atomic condensate with strong photoassociation
(compared to the s-wave collisional interaction); second, the photoassociation coupling is reduced, and a
macroscopic atom-molecule superposition is created via strong s-wave collisions; observation is based on a
Rabi-like oscillation of the system between the initial joint atom-molecule condensate and the corresponding
macroscopic superposition.  Here the main concern is to determine whether or not dissociative
decoherence prohibits the photoassociative formation of a macroscopic atom-molecule superposition.

Photoassociation occurs when two atoms absorb a photon, thereby jumping from the two-atom continuum to a
bound molecular state \cite{dr98,julienne98,we99, jm99,he00, kostrun00,ho01,va01,goral01,holland01,jm02}.
For initially quantum-degenerate atoms, there is an analogy in nonlinear optics: photoassociation is
formally identical to second-harmonic generated photons \cite{walls72}, while s-wave collisions between the
particles correspond to amplitude dispersion \cite{ys86}; in each case, the inherent non-linearity
entangles the particles, and potentially leads to macroscopic superpositions. In photoassociation, however,
the molecular state is excited and may spontaneously decay to molecular levels outside the system; also, the
molecule may dissociate either back to the atomic condensate, or to a pair of noncondensate atoms of equal
and opposite momentum via rogue \cite{jm99,kostrun00, jm02, mm03}, i.e., unwanted
\cite{goral01,holland01,NAI03}, photodissociation. These problems can be overcome by coupling the excited
molecular state with a stable molecular state using a second laser. In this two-color photoassociation, the
population loss via the excited molecular state can be made small by applying suitably large intermediate
detunings. Nevertheless, decay of the quantum coherence is still possible, even if the populations are
unaffected, and it is this possibility that warrents an invstigation of rogue dissociation in macroscopic
atom-molecule superpositions.

The possible creation of a macroscopic atom-molecule
superposition using photoassociation was reported in Calsamiglia et~al.~\cite{cms01}. Although lacking
decoherence effects, the results were promising and, moreover, Ref.
\cite{cms01} is an interesting starting point for our study for several other reasons: the possibility of a
superposition of atom-molecule condensates is tempting in itself, since atoms and molecules are different objects; the
whole dynamical process of creating the superposition state can be modelled; the
dissociation environment can be modelled in a straightforward manner. In Ref. \cite{hm05} the
dynamical approach has been used but, instead of photoassociation, the macroscopic superposition state was
created with the help of a magnetic field and Feshbach resonance. The main source of decoherence was the
laser-field induced interaction between atoms and the electromagnetic vacuum; however, the role of decoherence due to
the coupling between magnetic field modes and the condensate is unclear. Also, the study of Ref. \cite{cms01}
considers the
case of superposition very loosely. One cannot prove that the state is a superposition state by using
probability distributions only. Of course, the possibility of a Rabi-like oscillations to and from the
macroscopic superposition
\cite{cms01}, on which observation is based, is more likely in the presence of quantum coherence; but, close revivals
may also occur for nearly statistical mixtures. In all rigor, a detailed study of superposition states should
include the
off-diagonal elements of the density matrix, including the possibility of dissociative decoherence.

Hence, we focus on decoherence due to the second order term of the master
equation for the condensate modes in the perturbation theory (anomalous quantum correlations
induced by the first order term have been discussed elsewhere \cite{mm03}). Our results are roughly
summarized as follows. In general, the decoherence timescale $\tau_d$ should be longer than the particular
interaction timescale $\tau_i$, and thus macroscopic superpositions are possible only if the condition
\begin{equation}
\xi_d=\frac{\tau_d}{\tau_i}>>1
\end{equation}
is fulfilled. Our scheme is divided into two phases in which
different interactions are dominant. In phase $I$, we use photoassociation that is strong compared to
collisions to drive the initial atomic condensate into a particular joint atom-molecule which serves as an
initial state for the superposition engineering. Thus, the time scale $\tau_i$ is the photoassociation
timescale. In phase $II$, the intensity of photoassociation laser is turned down, and dominant collisions
between condensate particles drive the system into a macroscopic atom-molecule superposition. The
interaction timescale is then the collision timescale. With the given simulation parameters we can express
the coherence condition as
\begin{equation}
\xi_d=\frac{P_{I(II)}}{\rho[{\rm m^3}](2n^2+2n+1)},
\label{XID}
\end{equation}
where $P_I=2.712057\times 10^{10}$, $P_{II}=2.408246\times 10^{10}$, $\rho$ is the density
and $n$ the number of particles in the thermal cloud. The interesting fact is that the condition~(\ref{XID}) does
not depend on the number of particles in the initial condensate, $N$: With the chosen theoretical setup and
parameter values,
the photoassociation interaction, the collision interaction, and the coupling between molecular condensate and
noncondensate modes depend similarly on $N$; thus, we find that simulations for
$N=1000$ condensate are applicable to more realistic $N=10^8$ condensates. With values $\rho\sim 2\times
10^{19}~ {\rm m^{-3}}$ and $T\sim 10^{-9}~ {\rm K}$ we have $\xi_d\sim 3$, and the macroscopic atom-molecule
superposition will experience a considerable decay of off-diagonal elements. The best chance of avoiding rogue
decoherence when creating a macroscopic atom-molecule superposition is for 
$\rho=2\times 10^{18}\,{\rm m}^{-3}$ and $T\sim 0.1\,$nK, which gives $\xi_d\sim 10$, but which is just out of
reach of present ultracold technology~\cite{LEA03}.

The paper is outlined as follows. In Section \ref{theory} we sketch the photoassociation model given in Ref.
\cite{cms01} and the interactions with the uncorrelated thermal cloud (the environment). We calculate the
time evolution of the reduced density matrix, i.e., the master equation for the condensate modes only. Then,
in Sec. \ref{simulations} we assign explicit values to the model parameters and consider the structure of
our numerical analysis. The main results are also presented there. Section \ref{disc} is for summary and discussion.

\section{General theory \label{theory}}

\begin{figure}[b]
\includegraphics[width=8cm]{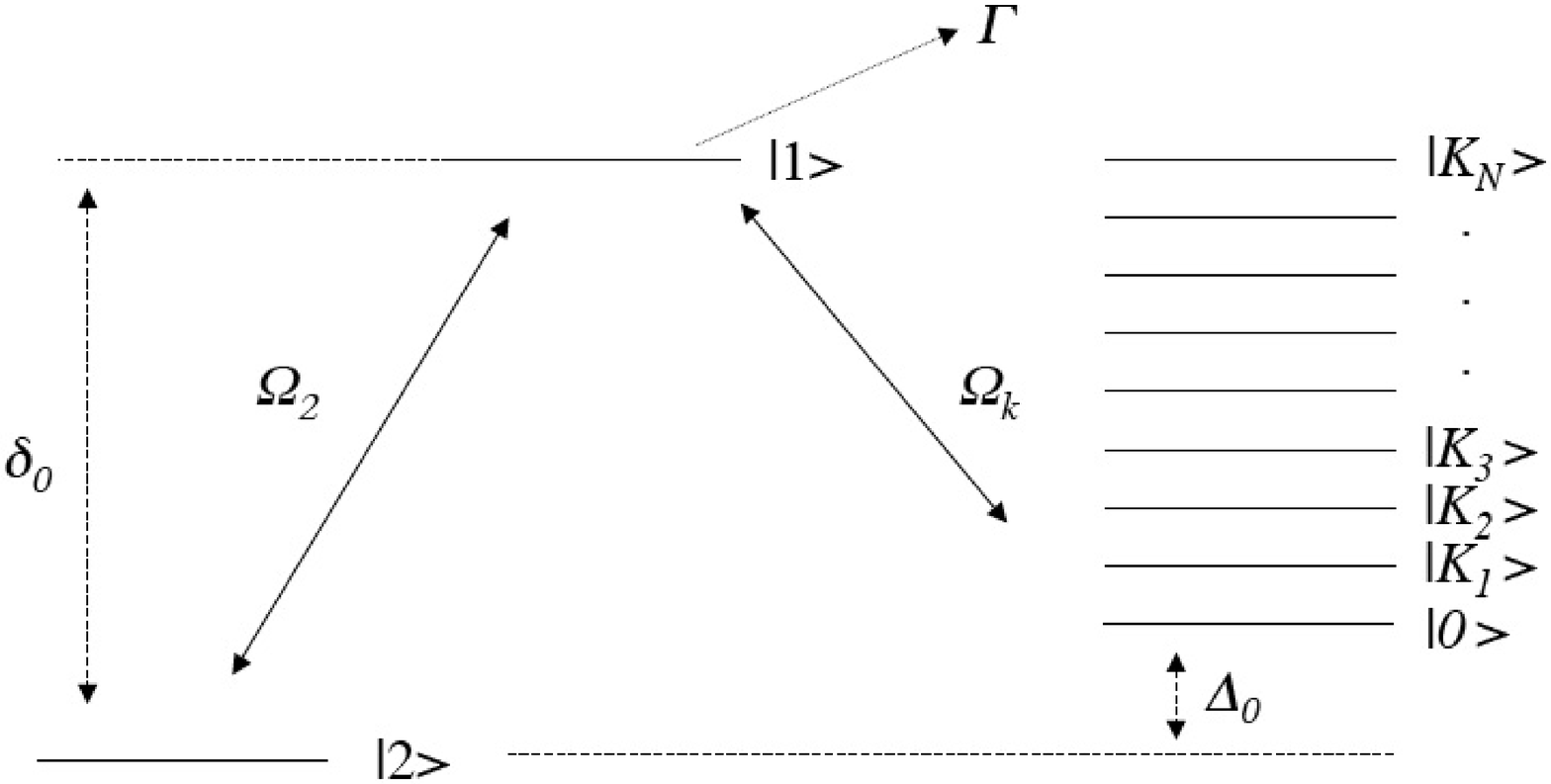}
\caption{\label{fig1} Few-level illustration of free-bound-bound
photoassociation. Initially, $N$ atoms are in the Bose-condensed state\
$\vert 0\rangle$. The free-bound laser then removes two atoms from this
state and creates an electronically-excited molecule in the state
$|1\rangle$. The bound-bound laser, in turn, removes an excited molecule
from the state $|1\rangle$, and creates an electronically-stable molecule
in state $|2\rangle$. The quasicontinuum of $N_{qc}$ noncondensate
disociation modes is also shown, where a pair of atoms with momentum
$\pm\hbar{\bf k}$ and energy $\hbar\epsilon_k=\hbar^2k^2/2m$ are taken to
occupy the state $|{\bf K}\rangle\equiv|{\bf k},-{\bf k}\rangle$.  The
free-bound and bound-bound Rabi frequencies are $\Omega_k=\Omega_1 f_{\bf k}$ 
and $\Omega_2$, where $f_k$ denotes the wavevector (energy)
dependence of the free-bound coupling. Lastly, the loss rate of the
electronically-excited molecular state is $\Gamma$, and the respective
intermediate and two-photon detunings are $\delta_0$ and
$\Delta_0$.}
\end{figure}

Consider a photoassociation laser that removes zero-momentum (${\bf k}=0$)
atoms from an initial atomic condensate $\vert 0\rangle$ and creates
molecules in the excited molecular condensate $\vert 1\rangle$. A second
laser couples this excited molecular condensate with a stable molecular
condensate $\vert 2\rangle$. The annihilation operators of the atomic
condensate, the excited molecular state and the stable molecular state
are denoted by $a_0\equiv a$, $b$ and $g$. Transitions to noncondensate
atoms arise because a molecule in $|1\rangle$ need not dissociate back to
the ${\bf k}=0$ atomic condensate $|0\rangle$, but may just as well wind
up as two atoms of equal and opposite momentum ($\pm{\bf k}$) in the
state $|{\bf K}\rangle\equiv|{\bf k},-{\bf k}\rangle$, since only
relative momentum is conserved. These noncondensate atoms are denoted by
$a_{\bf k}$, with $a_0\equiv a$ being the condensate. The interactions
that cause atom-molecule and molecule-molecule transitions are 
characterized by their respective Rabi frequencies $\Omega_{\bf
k}=\Omega_1 f_{\bf k}$ and $\Omega_2$, where $f_{\bf k}$ expresses the
wavevector dependence of the atom-molecule coupling. The one- and
two-photon detunings are $\Delta_0$ and
$\delta=\delta_0-i\frac{1}{2}\Gamma$, where the spontaneous decay rate, $\Gamma$, of the
excited molecular state is included in the one-photon
(intermediate) detuning. This scheme is illustrated in Fig. \ref{fig1}.
We also consider s-wave collisions between the atomic and 
stable-molecular condensates, with interaction strengths $\lambda_{aa}$,
$\lambda_{gg}$, and $\lambda_{ag}$. Anticipating large intermediate
detunings, the excited-state molecular fraction is low, and its
collisions can be ignored; similarly, we also neglect all collisions with
noncondensate atoms. Thus, in the rotating
wave approximation, the Hamiltonain for the above system is written as
\begin{eqnarray}
\frac{H_3}{\hbar} &=& -\Delta_0 g^{\dag}g + \delta b^{\dag}b +\sum_{\bf
k}\epsilon_{k} a^{\dag}_{\bf k} a_{\bf k} 
\nonumber\\&&
-\frac{1}{2}\sum_{\bf k}\left(\Omega_1f_{\bf k} 
b^{\dag}a_{\bf k}a_{-\bf k}+ h.c.\right)
-\left(\Omega_2 g^{\dag}b +h.c.\right)
\nonumber\\ &&
+2\lambda_{ag} a^{\dag}ag^{\dag}g 
+\lambda_{gg}g^{\dag}g^{\dag}gg+\lambda_{aa}a^{\dag}a^{\dag}aa.
\label{eq21}
\end{eqnarray}
For simplicity, we take $f_0=f^*_0=1$, $\Omega_j=\vert \Omega_j\vert
e^{i\phi_j}$, and $f_{\bf k}=f^*_{\bf k}$. We also assume free-particles,
so that $\epsilon_k=\hbar k^2/m$, where $m$ is the atomic mass. The
atomic condensate modes $a_0\equiv a$ are written explicitly in a moment.

We now develop an effective description by adiabatically eliminating
the excited molecular state from the Hamiltonian of Eq.~(\ref{eq21}),
based on the Heisenberg equations of motion
\begin{mathletters}
\begin{eqnarray}
i\dot{a} & = & -\vert\Omega_1\vert e^{i\phi_1} 
a^{\dag}b+2\lambda_{ag} ag^{\dag}g +2\lambda_{aa}a^{\dag}aa, 
\\ 
i\dot{a}_{\bf k} &=& \epsilon_{k}a_{\bf k}
-\frac{1}{2}\vert\Omega_1\vert e^{i\phi_1} f_{\bf k}a^{\dag}_{-\bf k}b,
\\
i\dot{b} & = & \delta b -\frac{1}{2}\vert\Omega_1\vert e^{i\phi_1} 
\sum_{\bf k}f_{\bf k}a_{\bf k}a_{-\bf k}
-\vert\Omega_2\vert e^{-i\phi_2}g,
\\
i\dot{g} & = & -\Delta_0 g -\vert\Omega_2\vert e^{i\phi_2} b
+2\lambda_{ag} a^{\dag}ag+ 2\lambda_{gg}g^{\dag}gg. \nonumber
\\
\end{eqnarray}
\end{mathletters}
The trick now is assuming that $\delta$ is the largest frequency in the
problem, and we can adiabatically eliminate the excited molecular state
by using $\dot{b}/\delta\sim 0$. Thus,
\begin{eqnarray}
b=\frac{\vert\Omega_1\vert e^{i\phi_1}}{2\delta} 
\sum_{\bf k}f_{\bf k}a_{\bf k}a_{-\bf k}
+\frac{\vert\Omega_2\vert e^{i\phi_2}}{\delta}g.
\end{eqnarray}
Inserting this into our three-level Hamiltonian of Eq. (\ref{eq21})
produces an effectively two-level Hamiltonian. The equations can be
simplified by denoting the relative phase of the lasers with
$\phi=\phi_2-\phi_1$, and by writing
$\chi=\vert\Omega_1\vert\vert\Omega_2\vert/\delta$. This yields 
\begin{eqnarray}
\frac{H_2}{\hbar} & = & -\Delta' g^{\dag}g 
+\sum_{\bf k}(\frac{1}{2}\Delta_0+\epsilon_{k})a^{\dag}_{\bf k}a_{\bf k}
\nonumber \\ &&
-\frac{\chi}{2}\sum_{\bf k}\left(e^{i\phi}g^{\dag}f_{\bf k}a_{\bf k} 
a_{-\bf k} + {\rm h.c.}\right)
\nonumber \\&&
+2\lambda_{ag}a^{\dag}ag^{\dag}g +\lambda'_{aa}a^{\dag}a^{\dag}aa
+\lambda_{gg}g^{\dag}g^{\dag}gg,
\label{eq22}
\end{eqnarray}
where $\Delta'=\Delta_0+ |\Omega_2|^2/\delta$ and
$\lambda'_{aa}=\lambda_{aa}-|\Omega_1|^2/4\delta$. Note that this is
exactly the same form of the Hamiltonian as for one-color transitions,
but the two-photon Rabi frequency $\chi$ has replaced the one-photon Rabi
frequency $\Omega_1$.

Next, we proceed to derive the master equation \cite{ms91,wm94} for the
stable molecular condensate, applying the same ideas as in Ref.
\cite{mm03}. Using the interaction picture, the problem is solved to
second order in perturbation theory. Taking the atom-molecule condensate
as the system, and the noncondensate modes as the environment, the
Hamiltonian of Eq. (\ref{eq22}) can be written as 
\begin{equation}
\frac{H_2}{\hbar}=\frac{H_S}{\hbar}+\frac{H_R}{\hbar}
-\left(\frac{\chi}{2} e^{i\phi}g^{\dag}
\sum_{{\bf k} \ne 0}f_{\bf k}a_{\bf k}a_{-\bf k}+ h.c.\right).
\label{eq1}
\end{equation}
The dynamics is not altered if, for later convenience, we add the constant
of motion $\lambda'_{aa}(N-N^2)$ to $H_2$ (see also Ref.~\cite{MAC05b}).
Thus, we get $H_S = H_0+H_I$, with
\begin{mathletters}
\begin{eqnarray}
\frac{H_0}{\hbar} & = & -\Delta g^{\dag}g, \\
\frac{H_I}{\hbar} & = & -\frac{1}2 \chi\left(e^{-i\phi}a^{\dag}a^{\dag}g 
+e^{i\phi}g^{\dag}aa\right)
\nonumber\\&&
-2\lambda g^{\dag}a^{\dag}ga+(\lambda_{gg}
-4\lambda'_{aa}) g^{\dag}gg^{\dag}g, \\
\frac{H_R}{\hbar} & = &\sum_{\bf k\ne 0}\epsilon_{k}a^{\dag}_{\bf k}
a_{\bf k},
\end{eqnarray}
\end{mathletters}
where $\Delta=\Delta'-2\lambda'_{aa}+\lambda_{gg}$, and
$\lambda=2\lambda'_{aa}-\lambda_{ag}$. We then apply the Born
approximation and calculate the master equation of our system in the
interaction picture. Initially the system $\vert n_m\rangle\vert
n_a\rangle$ and the environment $\prod_{\bf k\ne 0}\vert l_a\rangle_{\bf
k}$ are not correlated, i.e., $\rho_{tot} =\rho_{s} \otimes\rho_{e}$. The
equation of the motion of the total density matrix in the interaction
picture is 
\begin{equation}
\dot{\rho}_{tot}=-\frac{i}\hbar \left[H_{int} , \rho_{tot} \right].
\end{equation}

After integration, performing the trace over environment and neglecting
terms higher than second order, we get
\begin{eqnarray}
\dot{\rho}_{s} & = & -\frac{i}\hbar {\rm Tr}_{e}
\left[H_{int}(t),\rho_{s}(0)\otimes\rho_{e}\right]
\nonumber \\
& - & \frac{1}{\hbar^2}\int_0^t dt_1{\rm Tr}_{e} \left[H_{int}(t),
\left[H_{int}(t_1),\rho_{s}(0)\otimes\rho_{e}\right]\right]
\nonumber \\
& = & \dot{U}_1(t)+\dot{U}_2(t).
\label{eq2}
\end{eqnarray}
In the interaction picture, our Hamiltonian~(\ref{eq1}) becomes
\begin{equation}
H_{int}=-\hbar\left(A^{\dag} \Gamma+\Gamma^{\dag} A\right),
\end{equation}
where we have used the shorthand notation
\begin{mathletters}
\begin{eqnarray}
\Gamma & = & \sum_{{\bf k} \ne 0} f_{\bf k} a_{\bf k} a_{-\bf k} e^{-i(\epsilon_{k}-\Delta)t},
\\ A & = & \frac{\chi}{2}e^{-i\phi}g.
\end{eqnarray}
\end{mathletters}
Tracing of Eq. (\ref{eq2}) results in 
\begin{equation}
\dot{U}_1(t)=i\left(\langle \Gamma(t)\rangle \left[A^{\dag}, \rho_{s}(0)\right] 
+\langle\Gamma^{\dag}(t)\rangle \left[A,\rho_{s}(0)\right]\right)
\end{equation}
as the first order term, and the second order term is
\begin{widetext}
\begin{eqnarray}
\dot{U}_2(t) & = & -\int_0^t dt_1 \left( \langle \Gamma(t)\Gamma(t_1)\rangle
\left[A^{\dag},
\left[A^{\dag},\rho_{s}(0)\right]\right]+ \langle \Gamma^{\dag}(t)\Gamma(t_1)\rangle
\left[A,
\left[A^{\dag}, \rho_{s}(0)\right]\right] \right. \nonumber \\
 & + & \left. \langle \Gamma(t)\Gamma^{\dag}(t_1)\rangle \left[A^{\dag},
\left[A,\rho_{s}(0)\right]\right]+\langle \Gamma^{\dag}(t)\Gamma^{\dag}(t_1)\rangle
\left[A,
\left[A,\rho_{s}(0)\right]\right]\right).
\end{eqnarray}
\end{widetext}

The following six coefficients should be calculated:
\begin{mathletters}
\begin{eqnarray}
I_1 & = & \langle \Gamma(t)\rangle, \\
I_2 & = & \langle \Gamma^{\dag}(t)\rangle= I_1^*, \\
I_3 & = & \int_0^t dt_1  \langle \Gamma(t)\Gamma(t_1)\rangle, \\
I_4 & = & \int_0^t dt_1 \langle \Gamma^{\dag}(t)\Gamma^{\dag}(t_1)\rangle, \\
I_5 & = & \int_0^t dt_1 \langle \Gamma^{\dag}(t)\Gamma(t_1)\rangle, \\
I_6 & = & \int_0^t dt_1 \langle \Gamma(t)\Gamma^{\dag}(t_1)\rangle.
\end{eqnarray}
\end{mathletters}
The coefficients 
\begin{equation}
I_1=I_2=I_3=I_4=0,
\end{equation}
because for the normal thermalized heat bath the correlation function
$\langle a_{\bf k}a_{-\bf k}\rangle$ is zero, though these have been
studied elsewhere \cite{mm03}. In this study, the environment is a
thermal cloud around the condensates, and therefore, it is reasonable to
assume that it acts like an uncorrelated thermalized heat bath. The
nonzero correlations have the form of 
\begin{equation}
\langle a_{\bf k}^{\dag}a_{-\bf k}^{\dag}a_{\bf k'}a_{-\bf k'}\rangle = 4\pi n_k^2
\delta_{k,k'}.
\end{equation}

Let us calculate, say, $I_5$:
\begin{eqnarray}
I_5 & = & \int_0^t dt_1 \langle \Gamma^{\dag}(t)\Gamma(t_1)\rangle \nonumber
\\ & = & \int_0^t dt_1 \sum_{\bf k}\sum_{\bf k'}f_{\bf k}f_{\bf k'} 4\pi
n_k^2 \delta_{k,k'}
e^{i(\epsilon_k-\Delta)t}e^{-i(\epsilon_{k'}-\Delta)t_1} \nonumber \\ & = & 4\pi\int_0^t dt_1
\sum_k f_k^2n_k^2 e^{i(\epsilon_k-\Delta)(t-t_1)}.
\end{eqnarray}
The sum can be converted into an integral by assuming that the maximum
momentum is very large, i.e., infinite. This approximation will, however,
yield consequences that are not present in the original system (maximum
momentum very large but finite), e.g., irreversible dynamics. In our case
it means that the coherence of the system is decreasing monotonically.
Here we have assumed that the particles are in a free space. Of course,
if one wants to take the trap into account, one should add here the
density of the states of the trap. As the first approximation, we keep
the system as simple as possible; neglecting the trap can be justified by
the fact that the time scales of the trap are much longer than the
assumed decoherence time scale.

Next we go to the frequency representation: $k=\sqrt{m\epsilon/
\hbar}$, $d\epsilon=(2\hbar/m)kdk$, so that
\begin{equation}
I_5=\frac{m^{3/2}V}{\pi\hbar^{3/2}}\int_0^t dt_1 \int_0^\infty d\epsilon 
\sqrt{\epsilon}f^2(\epsilon)n^2(\epsilon)
e^{i(\epsilon-\Delta)(t-t_1)}.
\end{equation}
We assume that the Markov approximation is valid, i.e., $\Delta$ is large
enough and $\sqrt{\epsilon}f^2(\epsilon)n^2(\epsilon)$ is a slowly
varying function in the vicinity of $\epsilon=\Delta$, and thus the
correlation time scale is so short that we can set $t\rightarrow \infty$
in the integral $I_4$. These conditions are met when $\hbar\Delta\sim
kT$. This leads to the principal value integral
\begin{equation}
\lim_{t\rightarrow \infty}\int_0^t e^{i(\epsilon-\Delta)(t-t_1)}= \pi\delta(\epsilon-\Delta)
+i\frac{PV}{\Delta-\epsilon}.
\end{equation}
Thus,
\begin{equation}
I_5=\frac{m^{3/2}V}{\hbar^{3/2}}\sqrt{\Delta}f^2(\Delta) n^2(\Delta) +
i\Delta_*,
\end{equation}
where
\begin{equation}
\Delta_*= \frac{V m^{3/2}}{\pi\hbar^{3/2}}{\rm PV}\int_0^\infty d\epsilon 
\frac{\sqrt{\epsilon}}{\epsilon-\Delta}
f^2(\epsilon)e^{i(\epsilon-\Delta)(t-t_1)}n^2(\epsilon).
\end{equation}
While technically related to the Lamb shift, the term $\Delta_*$ is the shift that appears
anytime one connects a bound state to a continuum. This has been investigated in
theory~\cite{FED96,BOH99} and experiment~\cite{GER01,MCK02,PRO03}. The shift is
neglected in this study. Similarly, for $I_6$ we get 
\begin{equation}
I_6=\frac{m^{3/2}V}{\hbar^{3/2}}\sqrt{\Delta}f^2(\Delta) [n(\Delta)+1]^2.
\end{equation}

The master equation in the interaction picture is thus
\begin{eqnarray}
\dot{\rho}_{int} & = & -I_5\left[A, \left[A^{\dag},\rho\right]\right]
-I_6\left[A^{\dag},
\left[A,\rho\right]\right] \nonumber \\ & = & (I_5+I_6)(A\rho A^{\dag}+ A^{\dag}\rho A -
AA^{\dag}\rho -\rho A^{\dag}A). \nonumber\\
\end{eqnarray}
The corresponding master equation in the Schr\"odinger picture is
\begin{widetext}
\begin{eqnarray}
\dot{\rho}_{s} & = & -\frac{i}\hbar [H_0+H_I,\rho_{s}]+\dot{\rho}_{int} \nonumber
\\ & = & \frac{i}\hbar\left[\Delta (g^{\dag}g\rho-\rho g^{\dag}g)+
\frac{1}2\chi
e^{-i\phi}(a^{\dag}a^{\dag}g\rho-\rho a^{\dag}a^{\dag}g)+\frac{1}2\chi e^{i\phi}
(g^{\dag}aa\rho-
\rho g^{\dag}aa)\right. \nonumber \\ & - & \left.
(\lambda_{gg}-4\lambda'_{aa})(g^{\dag}gg^{\dag}g\rho-\rho
g^{\dag}gg^{\dag}g) +2\lambda(g^{\dag}a^{\dag}ga\rho- \rho g^{\dag}a^{\dag}ga)\right] \nonumber
\\ & + & (I_5+I_6)(A\rho A^{\dag}+ A^{\dag}\rho A - AA^{\dag}\rho -\rho
A^{\dag}A),
\end{eqnarray}
where $H_I$ is treated as a perturbation.
\end{widetext}

\section{Simulations \label{simulations}}
The aim of the present work is to study the impact of the rogue decoherence environment.
It is important to note that it is a matter of contention whether strong
two-color photoassociation of a Bose-Einstein condensate is even possible
\cite{MAC05}. We insist that it is and, moreover, that it is possible to
target ground-electronic state levels that are sufficiently low-lying to
allow neglect of vibrational-relaxation losses (see also \cite{MAC05b})
Our simulations therefore consist of two stages: first, a particular
joint-atom molecule condensate with real probability amplitude is created
in the regime where photoassociation is much stronger than collisions;
second, the photoassociation intensity is reduced and the joint
atom-molecule condensate evolves under now-dominant collisional
interaction into a macroscopic atom-molecule superposition.

Emipircally~\cite{cms01}, we have found that the parameter values that are sufficient to create the proper
Phase-$I$ joint atom-molecule condensate are:
\begin{mathletters}
\begin{eqnarray}
\Delta & = & 0, \\
\chi & = & 10\sqrt{N}\lambda, \\
\phi & = & \pi/2.
\end{eqnarray}
\end{mathletters}
By proper, we mean a joint atom-molecule condensate with a real
amplitude~\cite{cms01}, which is the origin of the relative phase
$\phi=\pi/2$, and where half the original atoms have been converted to molecules.
 The duration of the strong two-photon-resonant photoassociation pulse is $\tau$, 
which can be obtained from the solution of the semiclassical approximation for the 
molecules \cite{cms01}, $N/4=(N/2)\tanh^2(\sqrt{N}\chi\tau)$, which is a result 
borrowed from the theory of second-harmonic-generated photons~\cite{walls72}. The 
condition $\Delta=0$ means that the system is on a Stark-shifted resonance and the
Markov approximation becomes dubious. On resonance, the interaction
strength compared to the evolution time scale increases and one expects
that decoherence would take a more dominant role. Thus, we use a large
enough two-photon detuning, $\Delta>kT$, to avoid the resonance and
justify the Markov approximation: $\Delta=0.1N\lambda$, for
$\rho=1.88\times 10^{19} {\rm m^{-3}}$.

The second phase is collision-dominant:
\begin{mathletters}
\begin{eqnarray}
\Delta & = & \sqrt{N}\chi=0.1N\lambda, \\
\chi & = & 0.1\sqrt{N}\lambda, \\
\phi & = & 0.
\end{eqnarray}
\end{mathletters}
The collision interactions are
\begin{mathletters}
\begin{eqnarray}
\lambda_{aa} & = & \frac{4\pi\hbar a_{aa}}{mV}, \\
\lambda_{ag} & = & \frac{3\pi\hbar a_{ag}}{mV}, \\
\lambda_{gg} & = & \frac{2\pi\hbar a_{gg}}{mV}, 
\end{eqnarray}
\end{mathletters}
where the atom-atom scattering length is $a_{aa}=5.4$ nm
\cite{drummond02,wynar00}, the atom-molecule scattering length is
$a_{ag}=-9.346$ nm \cite{drummond02,wynar00}, the unknown
molecule-molecule scattering length is approximated as $a_{gg}=a_{aa}$,
the mass of a $^{87}{\rm Rb}$ atom is $m=1.443\times 10^{-25}$ kg, and
$V$ is the quantization volume that is expedient in a partiuclar context
(e.g., cubic box, spherical cavity, or harmonic trap). Also, by optically
tuning the atom-atom scattering
length~\cite{FED96,BOH97,BOH99,FAT00,THE04,THA05}, there is the
possibility to reduce the number of parameters by setting
$\lambda_{gg}-4\lambda'_{aa}=0$, which amounts to $a_{aa}'=a_{gg}/8$. The
effects of the loss term $\Gamma=12\times 2\pi~ {\rm MHz}$ \cite{wynar00}
should be negligible if $\vert\delta\vert >> \Gamma$; therefore, we
choose $\vert\delta\vert\sim10^3\times 2\pi~ {\rm MHz}$. For this
detuning, and barring an anomalously large molecule-molecule scattering
length, the light-induced scattering-length shift should be possible;
however, this really only a matter of convenience, and is not neccesary.
Finally, with the given value of $\lambda'_{aa}$, the remaining
collisional interaction strength is $\lambda=(\rho/N)\times 7.67715\times
10^{-17}\,{\rm Hz}$, where $[\rho]\equiv m^{-3}$.

We use the molecular number basis: $\rho=\vert m\rangle\langle m'\vert$, where $m$ and $m'$ are the molecule numbers,
and
$N$ is the total particle number. Hence, the coupled equations to solve are
\begin{widetext}
\begin{equation}
\begin{array}{cclcl}
\dot{\rho}_{m,m'} & = & \rho_{m,m'} \{i \Delta (m-m')+i 2 \lambda [(N-2m)m-(N-2m')m'] &-&
(I_5+I_6) \frac{\chi^2}{4} m'\}
\\ & - & \rho_{m,m'}(I_5+I_6)\frac{\chi^2}{4}(m+1) & ; & m<N/2
\\ & + & \rho_{m-1,m'}i\frac{1}{2}\chi e^{-i\phi}\sqrt{(N-2m+1)(N-2m+2)m}
& ; & m>0 \\
& - & \rho_{m,m'+1}i\frac{1}{2}\chi e^{-i\phi}\sqrt{(N-2m')(N-2m'-1)(m'+1)} & ; & m'<N/2 
\\ & + & \rho_{m+1,m'}i\frac{1}{2}\chi
e^{i\phi}\sqrt{(N-2m)(N-2m-1)(m+1)} & ; & m<N/2 \\
& - & \rho_{m,m'-1}i\frac{1}{2}\chi e^{i\phi}\sqrt{(N-2m'+1)(N-2m'+2)m'} & ; & m'>0
\\ & + & \rho_{m-1,m'-1}(I_5+I_6)\frac{\chi^2}{4}\sqrt{mm'} & ; & m,m'>0
\\
& + & \rho_{m+1,m'+1}(I_5+I_6)\frac{\chi^2}{4}\sqrt{(m+1)(m'+1)} & ; &
m,m'<N/2. 
\end{array}
\label{eq54}
\end{equation}
\end{widetext}
Let us stress that these equations of motion (\ref{eq54}) behave well at
the thermodynamical limit $V\rightarrow \infty$ because, as typical for
photoassociation, $\chi\sim 1/\sqrt{V}$ \cite{jm99, kostrun00},
$I_5+I_6\sim V$ and $\rho_{m,m'}\sim 1/N$.

The common decay profile of coherence in infinite models has the form of
$e^{-t/\tau_d}$, which defines the decoherence time $\tau_d$
\cite{uz89,zu91}. Phase $I$ follows this usual behavior, and has the
off-diagonal decoherence time
\begin{equation}
\tau_{d,{\rm off}}=\frac{4}{\chi^2(I_5+I_6)}=\frac{4\hbar^{3/2}\rho}
{m^{3/2}\chi^2\sqrt{\Delta}N(2n^2+2n+1)},
\label{DECOH_T}
\end{equation}
where the total particle number is $N$ and the fraction of noncondensate
atoms is $n=1/[\exp(\hbar\Delta/kT)-1]$. Since photoassociation is
dominant by design, this timescale must be compared to the
photoassociation time scale $\tau_{pa}=(\sqrt{N}\chi)^{-1}$. The
decoherence condition is thus 
\begin{equation} 
\xi_{d, I}=\frac{\tau_d}{\tau_{pa}}=\frac{2.712057\times 10^{10}}{\sqrt{\rho[{\rm
m^3}]}(2n^2+2n+1)},
\label{eq31}
\end{equation}
where we have used the detuning $\Delta=0.1N\lambda$. In most simulations
we have used a small but realistic temperature $T=10^{-9} {\rm K}$. Thus,
the validity of the Markov approximation requires $\Delta>131 {\rm Hz}$.
The nonzero detuning, compared with zero detuning, in the phase $I$ does
not contribute much to the moment $\tau$ when phase $I$ ends. The
generalized Rabi frequency \cite{har81} is $\chi_{gen} =
\sqrt{N\chi^2+\Delta^2} = N\lambda\sqrt{100+0.01}$. For simplicity, we do
not apply the time boundary between phases $I$ and $II$ given by the
semiclassical approximation [i.e., $\tau$ defined by
$N/4=(N/2)\tanh^2(\sqrt{N}\chi\tau)$], but we instead optimize the
distance between the two superposition peaks, or, the ``size" of the
superposition, to be a large as possible. An optimal superposition size,
and equally-sized probability peaks, are obtained only if the phase $I$
ends at the moment $\tau$ when ${\rm Tr}\rho g^{\dag}g\sim 0.296N/2$.
This end point is very sensitive: the accuracy in molecular fraction
should be $\sim \pm 0.001$ in order to get equally sized peaks. Also, the
end point depends a bit on the particle number. The above mentioned
result is valid for
$N=1000$.

Phase $II$ follows similar behavior, having the same decoherence time as
in Eq.~(\ref{DECOH_T}). However, since collisions are dominant in phase
$II$, we compare the decoherence time in phase $II$ with the collision
time scale $\tau_c\sim (2N\lambda)^{-1}$. Hence, the decoherence
condition is
\begin{equation}
\xi_{d, II}=\frac{\tau_d}{\tau_c}=\frac{2.408246\times 10^{10}}{\sqrt{\rho[{\rm
m^3}]}(2n^2+2n+1)}. 
\label{eq32}
\end{equation}
If phase $I$ results in suitable initial conditions to produce a
macroscopic superposition in phase $II$, decoherence in phase $II$ is the
only remaining thing to prevent the superposition state. The correlations
created by collisions are the mechanism that, in ideal conditions,
results in a macroscopic superposition of atomic and molecular
condensates. Therefore, if the decoherence time scale is smaller than the
collisional time scale then decoherence dominates over collisions and
prohibits the creation of the macroscopic superposition. The constraint
to the superposition here is $\xi_{d, II}>>1$.

\begin{figure}[t]

\includegraphics[width=8cm]{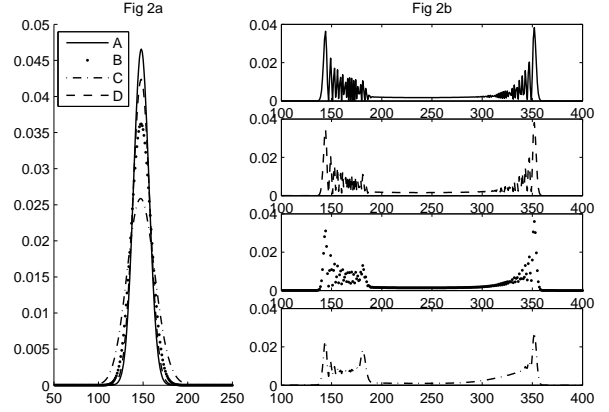}
\caption{The probability distributions of scenarios $A$ (no damping), $B$
(moderate damping), $C$ (strong damping) and $D$ (weak damping) at
(a)~$\lambda t_1=8.6\times10^{-5}$ when the stage $I$ ends and (b)~at
$\lambda t_2=1.0486\times10^{-2}$ when the superposition state emerges at
the end of stage $II$. Decoherence obviously affects the probability
distributions: stronger decoherence makes the probability peaks lower and
wider. Recall that the stage $I$ and $II$ decoherence conditions for the
various scenerios are ($A$)~$\xi_{d,I}=\xi_{d,II}\approx\infty$;
($B$)~$\xi_{d,I}=3.44$ and $\xi_{d,II}=3.05$;
($C$)~$\xi_{d,I}=\xi_{d,II}=1$; and,~($D$)$\xi_{d,I}=10.88$ and
$\xi_{d,II}=9.65$.}
\label{fig2} 
\end{figure}

\begin{figure}
\includegraphics[width=8cm]{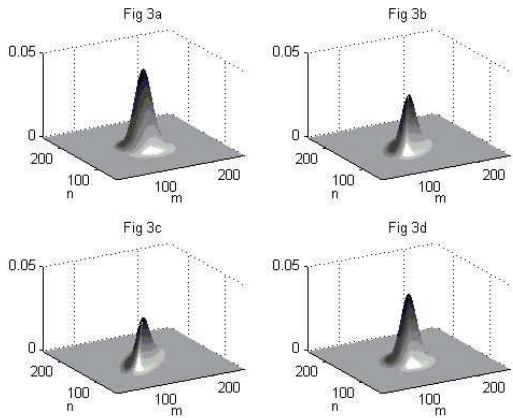}
\caption{\label{fig3} The density matrices of scenarios $A$~(a, no
damping), $B$~(b, moderate damping), $C$~(c, strong damping) and $D$~(d,
weak damping) at the end of stage $I$ at $t_1$. Decoherence effects are
clearly visible in scenarios $B$ and $C$, as the decay of off-diagonal
elements has resulted in ellipsoid-gaussian wave packet. Again, recall
the decoherence conditions: $A$, $\xi_{d,I}= \xi_{d,II}\sim\infty$; $B$,
$\xi_{d,I}=3.44$ and $\xi_{d,II}=3.05$; $C$ $\xi_{d,I}=\xi_{d,II}=1$; and
$D$, $\xi_{d,I}=10.88$ and
$\xi_{d,II}=9.65$.}
\end{figure}

\begin{figure}
\includegraphics[width=8cm]{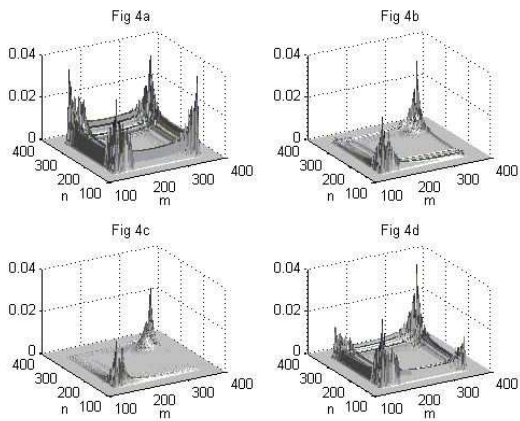}
\caption{\label{fig4} The density matrices of scenarios $A$~(a, no
damping), $B$~(b, moderate damping), $C$~(c, strong damping) and $D$~(d,
weak damping) at $t_2$, when a macroscopic superposition state should
emerge. Only a small fraction of off-diagonal elements have survived from
decoherence, even in the scenario $D$, that has extreme parameter values
of $T=10^{-10} {\rm K}$ and $\rho=10^{18} {\rm m^{-3}}$. Again, recall
the decoherence conditions: $A$, $\xi_{d,I}= \xi_{d,II}\sim\infty$; $B$,
$\xi_{d,I}=3.44$ and $\xi_{d,II}=3.05$; $C$ $\xi_{d,I}=\xi_{d,II}=1$; and
$D$, $\xi_{d,I}=10.88$ and
$\xi_{d,II}=9.65$.}
\end{figure}

There are several damping scenerios to consider; although, due to limited
computational resources, we consider the case of $N=1000$ particles. In
scenario $A$ there is no decoherence present, i.e.,
$\xi_{d,I}=\xi_{d,II}\approx\infty$. In scenerio $B$, the parameter
values $\rho=1.88\times 10^{19}~ {\rm m^{-3}}$ and $T=10^{-9}~{\rm K}$
result in moderate decoherence, i.e., $\xi_{d,I}=3.44$ and
$\xi_{d,II}=3.05$. The scenario $C$ is the borderline scenario with
$\xi_{d,I}= \xi_{d,II}=1$. Corresponding temperatures and densities can
be calculated from Eqs. (\ref{eq31}) and (\ref{eq32}). In scenario $D$,
we assume a temperature $T=10^{-10}~ {\rm K}$ and density
$\rho=1.88\times 10^{18} {\rm m^{-3}}$, i.e., $\xi_{d,I}=10.88$ and
$\xi_{d,II}=9.65$. The results are illustrated in
Figs.~\ref{fig2}-\ref{fig4}. In Fig.~\ref{fig2} we present the
probability distributions of all scenarios at two different instants of
time. The first instant ($t_1$) is when the phase $I$ ends, and a
particular joint-atom molecule condensate with a real probability
amplitude is created. It serves most ideal initial conditions to begin
the collision dominant phase $II$. The second instant ($t_2$) is when the
superposition peaks should emerge. Figure~\ref{fig3} presents the density
matrices of all scenarios at $t_1$, and Fig. \ref{fig4} the density
matrices of all scenarios at $t_2$.

Overall, while decoherence is strong enough, the quantum correlations are
damped out and sharp peaks in probability distributions smoothen. The
decay of off-diagonal elements seems to be considerable also in the
regime of moderate decoherence $1<\xi_d<6$. Strong decoherence effects
$\xi_d<1$ would be even faster. In order to get a macroscopic
superposition with the given setup, one should be safely in the regime of
weak decoherence, i.e., $\xi_d>10$. However, the regime is not reachable
using parameter values consistent with present-day technology. Of course,
there are a few possibilities to improve the result and get $\xi_d>10$.
First, one can reduce the density $\rho$, but $10^{19} {\rm m^{-3}}$ is
somewhat rarified as it is, even for a condensate, and achieving $2\times
10^{18} {\rm m^{-3}}$ could present challenges. The second possibility is
to enter the non-Markovian regime, which could considerably improve the
$\xi_d$ at least in phase $I$; however, the non-Markovian case will be
studied elsewhere. Other tricks are related to manipulating the
interaction or the environment. The symmetrization of the environment
\cite{dalvit00} would be a hard, maybe impossible, task to apply here due
to the inherent nonlinearity; but, creating specially a correlated
environment might help since an uncorrelated heat bath is quite a harsh
environment for coherence phenomena.

Finally, we note that the observation of the possible superposition state
could present a problem. The assumption that the coherence of the state after
phase $I$ can be verified, and thus one only needs to consider the
density matrix profiles of the revival of this initial state after the
superposition state \cite{cms01,hm05}, is perhaps too simple, at least in
the Markovian regime. While the system is driven from superposition to
the revived state, rogue dissociation is still damping coherence out of
the system. Also, coherent tricks, such as phase-imprinting the revived
state, would need extra time, and this introduces more decoherence. The
best candidate for observing the macroscopic superposition is probably to
try to entangle the superposition state with another system, and then to
apply Bell-type correlation experiments.

\section{Discussion \label{disc}}
The important point is that our approach is a dynamical one in which we
{\it make the macroscopic superposition in the presence of the
decoherence}, which differs from the standard way in which one takes the
macroscopic superposition for granted and then puts it into a decoherent
environment and looks how long it would last there. Such studies do not
make claims whether it is possible to create a macroscopic superposition
or not, but they only make claims whether the initially prepared
macroscopic superposition can be observed or not. Therefore, our
photoassociation-based approach suggests a possible experimental {\it
creation} of the macroscopic superposition in a decohering environment.

Hence, the aim of this study is to understand the effects of the
rogue-dissociation-induced decoherence while trying to create a
macroscopic superposition of Bose-condensed atoms and molecules. In
particular, ours is a two-stage scheme: in the first stage, the
photoassociation coupling is strong compared to the s-wave collisional
coupling, and a specific joint atom-molecule condensate is created; in
the second stage, the photoassociation coupling is turned down such that
it is weak compared to the collisional coupling, and the system evolves
into a macroscopic atom-molecule superposition. There is a constraint
$\xi_d>>1$, which should be satisfied in order to get a macroscopic
superposition of atoms and molecules. The dependency on relevant system
parameters of the constraint $\xi_d$ was calculated in both stages of our
theoretical setup. An interesting fact is that, for the given parameter 
ratios, $\xi_d$ does not depend on the number of particles in condensate
$N$. Results for a modest, say, $N=1000$ particle superposition state
will therefore apply to a $N=10^9$ particle superposition state with the
same density, given properly scaled laser frequencies and intensities.
Using $\xi_d$ one can easily, without heavy simulations, consider whether
the macroscopic superposition state of atoms and molecules is within
reach or not with given experimental setup parameters.

Without any tricks, the best chance to defeat rogue decoherence in
creating an atom-molecule superposition corresponds to a density 
$\rho=2\times 10^{18}~ {\rm m^{-3}}$ and temperature $T\sim 0.1~ {\rm nK}$. 
Nevethless, while this temperature may be accessible by making the trap 
shallow, the off-diagonal elements of density matrix will still experience
considerable decay. One way to get better macroscopic superpositions in
the present scheme is to challenge the assumptions or approximations of
our study or, in other words, engineer the environment. We assumed that
the environment is an uncorrelated heat bath, the system parameter values
should be in the regime where Markov approximation is valid, the
environment particles are free and the environment is infinite. Indeed, it
may well prove interesting to examine the non-Markovian regime, where a
greater part of the physically-relevant parameter space is available,
and/or to use a specially correlated environment.

Rogue dissociation is an important source of
decoherence that should not be neglected in coherence studies of atom-molecule
Bose-Einstein condensates. Unfortunately, it seems to push coherent atom-molecule
engineering (macroscopic superpositions, quantum computation, etc.) one
step further away from realization. 

\acknowledgments
The authors gratefully acknowledge Kalle-Antti Suominen for helpful
discussions, financial support from the Academy of Finland, project
206108 and Magnus Ehrnrooth Foundation (OD), and the National Science Foundation (MM).


\begin{thebibliography}{99}

   \bibitem[Schr\"odinger (1935)]{sc35}
    E. Schr\"odinger, Die Naturwissenschaften {\textbf{23}}, 807-812, 823-828, 844-849 (1935). 

   \bibitem[Zurek (1982)]{zu82}
    W. H. Zurek, Phys. Rev. D {\textbf{26}}, 1862 (1982). 

   \bibitem[Caldeira \& Leggett (1983)]{cl83}
    A. O. Caldeira and A. J. Leggett,
    Physica A {\textbf{121}}, 587 (1983);
    Phys. Rev. A {\textbf{31}}, 1059 (1985). 

   \bibitem[Unruh et Zurek (1989)]{uz89}
    W. G. Unruh and W. H. Zurek,
    Phys. Rev. D {\textbf{40}}, 1071 (1989).

   \bibitem[Zurek (1991)]{zu91}
    W. H. Zurek
    Phys. Today {\textbf{44}}(10), 36 (1991).

   \bibitem[Hu et al. (1992)]{hpz92}
    B. L. Hu, J. P. Paz, and Y. Zhang,
    Phys. Rev. D {\textbf{45}}, 2843 (1992).

   \bibitem[Anglin et al. (1995)]{alzp95}
    J. R. Anglin, R. Laflamme, W. H. Zurek, and J. P. Paz,
    Phys. Rev. D {\textbf{52}}, 2221 (1995).

   \bibitem[Brune et al. (1996)]{br96}
    M. Brune, E. Hagley, J. Dreyer, X. Ma\^{i}tre, A. Maali,
    C. Wunderlich, J. M. Raimond, and S. Haroche,
    Phys. Rev. Lett. {\textbf{77}}, 4887 (1996). 

   \bibitem[Ruostekoski et al. (1998)]{ru98}
    J. Ruostekoski, M. J. Collett, R. Graham, and D. F. Walls,
    Phys. Rev. A {\textbf{57}}, 511 (1998).

   \bibitem[Cirac et al. (1998)]{ci98}
    J. I. Cirac, M. Lewenstein, K. M\o lmer, and P. Zoller,
    Phys. Rev. A {\textbf{57}}, 1208 (1998).

   \bibitem[Gordon and Savage (1999)]{cs99}
    D. Gordon and C. M. Savage,
    Phys. Rev. A {\textbf{59}}, 4623 (1999). 

   \bibitem[Dalvit, Dziarmaga, and Zurek (2000)]{dalvit00}
    D. A. R. Dalvit, J. Dziarmaga, and W. H. Zurek,
    Phys. Rev. A {\textbf{62}}, 013607 (2000).

   \bibitem[Calsamiglia, Mackie \& Suominen (2001)]{cms01}
    J. Calsamiglia, M. Mackie, and K.-A. Suominen,
    Phys. Rev. Lett. {\textbf{87}}, 160403 (2001).

   \bibitem[Huang et Moore (2005)]{hm05}
    Y. P. Huang, and M. G. Moore,
    /cond-mat/0508659 (2005).   

   \bibitem[Drummond, Kheruntsyan, and He (1998)]{dr98}
    P. D. Drummond, K. V. Kheruntsyan, and H. He,
    Phys. Rev. Lett. {\textbf{81}}, 3055 (1998).

   \bibitem[Julienne et al. (1998)]{julienne98}
    P. S. Julienne, K. Burnett, Y. B. Band, and W. C. Stwalley,    
    Phys. Rev. A {\textbf{58}}, R797 (1998).

   \bibitem[Weiner et al. (1999)]{we99}
    J. Weiner, V. S. Bagnato, S. Zilio, and P. S. Julienne,
    Rev. Mod. Phys. {\textbf{71}}, 1 (1999).

   \bibitem[Javanainen et Mackie (1999)]{jm99}
    J. Javanainen, and M. Mackie,
    Phys. Rev. A {\textbf{59}}, R3186 (1999).

   \bibitem[Kostrun et al. (2000)]{kostrun00}
    M. Ko\v{s}trun, M. Mackie, R. C\^{o}t\'{e}, and J. Javanainen,
    Phys. Rev. A {\textbf{62}}, 063616 (2000).

   \bibitem[Heinzen et al. (2000)]{he00}
    D. J. Heinzen, R. Wynar, P. D. Drummond, and K. V. Kheruntsyan,
    Phys. Rev. Lett. {\textbf{84}}, 5029 (2000).

   \bibitem[Hope et Olsen (2001)]{ho01}
    J. J. Hope and M. K. Olsen,
    Phys. Rev. Lett. {\textbf{86}}, 3220 (2001).

   \bibitem[Vardi et al. (2001)]{va01}
    A. Vardi, V. A. Yurovsky, and J. R. Anglin,
    Phys. Rev. A {\textbf{64}}, 063611 (2001).

   \bibitem[Javanainen et Mackie (2002)]{jm02}
    J. Javanainen, and M. Mackie,
    Phys. Rev. Lett. {\textbf{88}}, 090403 (2002).

   \bibitem[Goral et al. (2001)]{goral01}
    K. G\'{o}ral, M. Gajda, and K. Rz\c{a}\.{z}ewski,
    Phys. Rev. Lett. {\textbf{86}}, 1397 (2001).

   \bibitem[Holland et al. (2001)]{holland01}
    M. Holland, J. Park, and R. Walser,
    Phys. Rev. Lett. {\textbf{86}}, 1915 (2001).

   \bibitem{NAI03}
    P. Naidon and F. Masnou-Seeuws, 
    \pra {\bf 68}, 033612 (2003).

   \bibitem[Walls et Tindle (1972)]{walls72}
    D. F. Walls and C. T. Tindle,
    J. Phys. A {\textbf{5}}, 534 (1972).

   \bibitem[Yurke et Stoler (1986)]{ys86}
    B. Yurke and D. Stoler,
    Phys. Rev. Lett. {\textbf{57}}, 13 (1986);
    Phys. Rev. A {\textbf{35}}, 4846 (1987). 

   \bibitem[Mackie (2003)]{mm03}
    M. Mackie,
    Phys. Rev. Lett. {\textbf{91}}, 173004 (2003).

   \bibitem{LEA03}
    A. E. Leanhardt, T. A. Pasquini, M. Saba, A. Schirotzek, Y. Shin, D. Kielpinski, D. E. Pritchard, 
    and W. Ketterle, Science {\bf 301}, 1513 (2003).

   \bibitem[Meystre \& Sargent III (1991)]{ms91}
    P. Meystre and M. Sargent III,
    Elements of Quantum Optics (2nd ed.),
    Springer-Verlag, Heidelberg (1991).

   \bibitem[Walls \& Milburn (1994)]{wm94}
    D. F. Walls and G. J. Milburn,
    Quantum Optics,
    Springer-Verlag, Heidelberg (1994).

   \bibitem{MAC05b}
    M. Mackie, K. H\"ark\"onen, A. Collin, K.-A. Suominen, and J. Javanainen,
    \pra {\bf 70}, 013614 (2004).

   \bibitem{FED96}
    P. O. Fedichev, Yu. Kagan, G. V. Shlyapnikov, and J. T. M. Walraven,
    \prl {\bf 77}, 2913 (1996).

   \bibitem{BOH99}
    J. L. Bohn and P. S. Julienne,
    \pra {\bf 60}, 414 (1999).

   \bibitem{GER01}
    J. M. Gerton, B. J. Frew, and R.G. Hulet,
    \pra {\bf 64}, 053410 (2001).

   \bibitem{MCK02}
    C. McKenzie {\it et al.},
    \prl {\bf 88}, 120403 (2002).

   \bibitem{PRO03}
    I. D. Prodan, M. Pichler, M. Junker, and R. G. Hulet,
    \prl {\bf 91}, 080402 (2003).

   \bibitem{MAC05}
    M. Mackie, A. Collin, and J. Javanainen,
    \pra {\bf 71}, 017601 (2005);
    P. D. Drummond, K. V. Kherunstyan, D. J. Heinzen, and R. Wyner,
    \pra {\bf 71}, 017602 (2005).

   \bibitem[Drummond et al. (2002)]{drummond02}
    P. D. Drummond, K. V. Kheruntsyan, D. J. Heinzen, and R. H. Wynar, 
    Phys. Rev. A {\textbf{65}}, 063619 (2002).

   \bibitem[Wynar et al. (2000)]{wynar00}
    R. H. Wynar, R. S. Freeland, D. J. Han, C. Ryu, and D. J. Heinzen, 
    Science {\textbf{287}}, 1016 (2000).

   \bibitem{BOH97}
    J. L. Bohn and P. S. Julienne,
    \pra {\bf 56}, 1486 (1997).

   \bibitem{FAT00}
    F. K. Fatemi, K. M. Jones, and P. D. Lett,
    \prl {\bf 85}, 4462 (2000).

   \bibitem{THE04}
    M. Theis, G. Thalhammer, K. Winkler, M. Hellwig, G. Ruff, R. Grimm, and J. Hecker Denschlag,
    \prl {\bf 93}, 123001 (2004).

   \bibitem{THA05}
    G. Thalhammer, M. Theis, K. Winkler, R. Grimm, and J. Hecker-Denschlag,
    \pra {\bf 71}, 033403 (2005).

   \bibitem[Harter et al. (1981)]{har81}
    D. J. Harter, P. Narum, M. G. Raymer, and R. W. Boyd,
    Phys. Rev. Lett. {\textbf{46}}, 1192-1195 (1981).

\end{thebibliography}
\end{document}